\documentclass[12pt,floatfix,showkeys,superscriptaddress,aps,prd,preprint]{revtex4}
\usepackage[latin1]{inputenc}
\usepackage[T1]{fontenc}
\usepackage{lmodern}
\usepackage{amsmath}
\usepackage{amssymb}
\usepackage{graphicx}
\usepackage{esint}
\usepackage{longtable}
\usepackage{dcolumn}
\usepackage[english]{babel}
\usepackage{csquotes}
\usepackage{color}
\usepackage{slashed}
\usepackage{simplewick}
\usepackage{tikz-feynman}
\usepackage[]{tikz-feynman}
\usepackage{amsmath,latexsym}
\usepackage{color}

\usepackage{hyperref}
\hypersetup{
    colorlinks,
    citecolor=blue,
    filecolor=green,
    linkcolor=purple,
    urlcolor=red,
}

\usepackage{slashed}

\usepackage{hyperref}
\hypersetup{colorlinks,breaklinks,
			citecolor=[rgb]{0,0.0,1.0},
            urlcolor=[rgb]{0.0,0.0,1.0},
            linkcolor=[rgb]{0,0.5,0.9}}

\def\be{\begin{equation}}
\def\ee{\end{equation}}
\def\bea{\begin{eqnarray}}
\def\eea{\end{eqnarray}}

\def\d#1#2{\frac{\displaystyle #1}{\displaystyle #2}}

\def\m{\mu}
\def\n{\nu}

\begin{document}

\title{
A new class of regular black hole solutions with quasi-localized sources of matter in $(2 + 1)$ dimensions
}

%%%%%%%%%%%%%%%%%%%%%%
%%%%%%%%%%%%%%%%%%%%%%%%%%%%%%%%%%%%%%%%%%%%%%%%%%%%%%%%%%%%%%%%%%%%%%
\author{R. V. Maluf}
\email{r.v.maluf@fisica.ufc.br}
\affiliation{Universidade Federal do Cear\'a (UFC), Departamento de F\'isica,\\ Campus do Pici, Fortaleza - CE, C.P. 6030, 60455-760 - Brazil.}

%%%%%%%%%%%%%%%%%%%%%%%%%%%%%%%%%%%%%%%%%%%%%%%%%%%%%%%%%%%%%%%%%%%%%%%
\author{C. R. Muniz}
\email{celio.muniz@uece.br}
\affiliation{Universidade Estadual do Cear\'a (UECE), Faculdade de Educa\c{c}\~ao, Ci\^encias e Letras de Iguatu, Av. D\'ario Rabelo s/n, Iguatu-CE, 63.500-00 - Brazil.}

%%%%%%%%%%%%%%%%%%%%%%%%%%%%%%%%%%%%%%%%%%%%%%%%%%%%%%%%%%%%%%%%%%%%%%
\author{A. C. L. Santos}
\email{alanasantos@fisica.ufc.br}
\affiliation{Universidade Federal do Cear\'a (UFC), Departamento de F\'isica,\\ Campus do Pici, Fortaleza - CE, C.P. 6030, 60455-760 - Brazil.}

%%%%%%%%%%%%%%%%%%%%%%%%%%%%%%%%%%%%%%%%%%%%%%%%%%%%%%%%%%%%%%%%%%%%%%

\author{Milko Estrada}
\email{milko.estrada@gmail.com}
\affiliation{Facultad de Ingenier\'ia, Ciencia y Tecnolog\'ia, Universidad Bernardo O'Higgins, Santiago, Chile.}

\date{\today}

\begin{abstract}
This paper investigates a new class of regular black hole solutions in (2 + 1)-dimensions by introducing a generalization of the quasi-localized matter model proposed by Estrada and Tello-Ortiz. Initially, we try to physically interpret the matter source encoded in the energy-momentum tensor as originating from nonlinear electrodynamics. We show, however, that the required conditions for the quasi-locality of the energy density are incompatible with the expected behavior of nonlinear electrodynamics, which must tend to Maxwell's theory on the asymptotic limit. Despite this, we propose a generalization for the quasi-localized energy density that encompasses the existing models in the literature and allows us to obtain a class of regular black hole solutions exhibiting remarkable features on the event horizons and their thermodynamic properties. Furthermore, since the usual version of the first law of thermodynamics, due to the presence of the matter fields, leads to incorrect values of entropy and thermodynamics volume for regular black holes, we propose a new version of the first law for regular black holes.

\end{abstract}

%\pacs{04.70.-s,04.50.Kd,11.30.Cp,04.60.-m}
\keywords{regular black hole, ~nonlinear electrodynamics, ~$(2+1)$-dimensional spacetime}

\maketitle

%%%%%%%%%%%%%%%%%%%%%%%%%%%%%%%%%%%%%%%%%%%%%%%%%%%%%%%%%%%%%%%%%%%%%%%%%%%%%%%%%%%%%%

\section{Introduction}

Some solutions of gravitational field equations in general relativity (GR) have divergences (singularities), as is the case with the usual black hole solutions. These singularities could be related to a limitation in a classical theory and would be corrected in a quantum version \cite{Bambi:2013ufa}. In 1968, yet inside a classical description, James Bardeen \cite{BARDEEN} found the first regular black hole solution in (3 + 1)-dimensions. In other words, a solution without divergences in the metric and their curvature invariants. Afterward, Ay\'{o}n-Beato and Garc\'{i}a \cite{Ayon-Beato:2000mjt} proposed the interesting interpretation that the metric chosen by Bardeen is equivalent to taking nonlinear electrodynamics as the source of the energy-momentum tensor. Since then, these solutions for regular black holes have been widely explored in literature \cite{Balart:2014cga, Burinskii:2002pz, Bronnikov:2000vy, Ma:2015gpa, Toshmatov:2017zpr}, and recent works include quantum corrections \cite{Sharif:2010pj, Maluf:2018ksj}, thermodynamics analysis \cite{Akbar}, cosmological constant, and quintessence \cite{Rodrigues:2022qdp,Celio}. For details about the issue involving nonlinear electrodynamics as a matter source, one can refer to the review article \cite{Sorokin:2021tge}.

There are also regular solutions in lower-dimensional spacetime \cite{Cataldo:2000ns,He:2017ujy}, and their studies have attracted attention for the special properties which they exhibit. In particular, (2 + 1)-dimensional gravity models work as a simplest theoretical laboratory than its higher-dimensional counterpart, bringing new insights to fundamental questions of classical and quantum gravity \cite{Carlip:1995qv}. In (2+1)-dimensions, the best-known black hole solution was discovered in the early 1990s by Ba\~{n}ados, Teitelboim, and Zanelli (BTZ) \cite{Banados:1992wn,Banados:1992gq}. It is known that in (2+1)-dimensions, general relativity does not have a Newtonian limit in the weak field regime, and the gravitational field has no dynamical degrees of freedom \cite{Gott:1982qg,Barrow1986}. In this case, the curvature is determined by the Ricci tensor, and in the absence of matter, we have $R_{\mu\nu}=0$, so we cannot have a black hole solution \cite{Padmanabhan:2010zzb}. To avoid this problem, a negative cosmological constant was introduced into the theory allowing black hole solutions to existing in a spacetime asymptotically anti-de Sitter (AdS). Since the BTZ solution was found, several other (2+1)-dimensional exact black hole solutions have been obtained in different scenarios \cite{Garcia-Diaz:2017cpv}.

Regular black hole solutions have been proposed in (2 + 1)-dimensional spacetime \cite{Bueno:2021krl}, considering a profile of mass density that obeys the quasi-locality of the corresponding mass-energy \cite{Estrada:2020tbz, Hendi:2022opt}. In this context, Estrada and Tello-Ortiz \cite{Estrada:2020tbz} proposed a mass density that satisfies the requirements of a position-dependent mass function that is quasi-localized, i.e., the mass function must reach its finite maximum value at infinity. Inspired by this proposal, we seek to interpret the origin of the matter source as coming from nonlinear electrodynamics. As a result, we show that the quasi-locality requirements and Maxwell's theory limit for nonlinear electrodynamics are incompatible. Despite this result, we propose a generalization for the mass density function that gives rise to a new class of regular black hole solutions.

On the other hand, the standard version of the first law of thermodynamics must be modified for regular black holes due to the presence of matter fields in the energy-momentum tensor \cite{Ma:2014qma}. In fact, the usual first law, namely, $dm=TdS-PdV$, 
leads to entropy and thermodynamics volume values that do not coincide with the Bekenstein-Hawking area law \cite{Bekenstein:1973ur}. Aiming to fix this problem, Ma and Zhao included an additional factor in the first law, which corresponds to an integration of the radial coordinate up to infinity \cite{Ma:2014qma}. In Ref. \cite{Estrada:2020tbz}, a local definition of energy at the horizon was proposed,  imposing constraints on the evolution along the space parameters. Thus, including matter fields requires redefining the energy term to get the correct values for temperature, entropy, and volume, so the first law of thermodynamics for regular black holes is still an open question. In this context, we propose a new version of the first law for regular black holes based on the inclusion of a potential $dX$, which leads to a local definition of internal energy at the horizon, $dU=dm+dX$. So, we study the thermodynamic properties of these new solutions, calculating Hawking temperature, entropy, Gibbs potential, and heat capacity.

This paper is organized as follows: In section \ref{review}, we determine the mapping among perfect fluids densities and nonlinear electrodynamics. We proceed with the thermodynamics analysis in section \ref{thermal}. Finally, in section \ref{conclusion}, we present our remarks and conclusions.

%%%%%%%%%%%%%%%%%%%%%%%%%%%%%%%%%%%%%%%%%%%%%%%%%%%%%%%%%%%%%%%%%%%%%%%%%%%%%%%%%%%%%%%%%%
\section{(2 + 1)-Regular black hole with nonlinear electrodynamics
\label{review}}
The (2 + 1)-Einstein gravity action coupled to nonlinear electrodynamics can be defined as
\be\label{action}
S=\int d^3x\sqrt{-g}\left[\frac{R-2\Lambda}{16\pi}+L(F)\right],
\ee
where $g$ is the determinant of the metric tensor, $\Lambda=-1/\ell^2$ is the cosmological constant written in terms of the anti-de Sitter (AdS) radius, $\ell$, and $L(F)$ is the Lagrangian of the nonlinear electrodynamics with $F=F^{\mu\nu}F_{\mu\nu}$ \cite{Cataldo:2000ns,He:2017ujy}.

The Einstein field equations are obtained by varying the action (\ref{action}) with respect to the metric tensor, yielding,
\be\label{GReq}
G_{\mu\nu}+\Lambda g_{\mu\nu}=8\pi T_{\mu\nu},
\ee
with the energy momentum tensor given by
\be
T_{\mu\nu} \equiv -\d{2}{\sqrt{-g}}\d{\delta(\sqrt{-g} L_m)}{\delta g^{\mu\nu}}=g_{\mu\nu}L(F)-4L_{,F}F_{\mu\alpha}F_{\nu}^{~\alpha},
\ee
where $L_{,F}$ represents the derivative of $L(F)$ with respect to $F$. The electromagnetic field equations are determined by varying with respect to the potential $A_{\mu}$, remembering that the field-strength tensor $F_{\mu\nu}$ is defined as $F_{\mu\nu}= \partial_{\mu}A_{\nu}-\partial_{\nu}A_{\mu}$, we obtain
\be\label{EMeq}
\nabla_\mu(L_{,F}F^{\mu\nu})=0.
\ee
To find exact solutions to the above field equations, let us assume a static, circularly symmetric spacetime whose metric ansatz is given by
\be\label{metric}
ds^2=-f(r)dt^2+f(r)^{-1}dr^2+r^2d\phi^2,
\ee where $f(r)$ is an arbitrary function of the radial coordinate $r$.
Furthermore, it was demonstrated in Ref.  \cite{Cataldo:2000ns} that the magnetic field vanishes and only the electric field plays the role of source of the gravitational field in a geometry defined by the metric ansatz (\ref{metric}). Therefore, we can assume that the electromagnetic field tensor is expressed in the simplified form
 \be
 F_{\mu\nu}=E(r)(\delta^t_{\m}\delta^r_{\n}- \delta^t_{\n}\delta^r_{\m}),\label{F}
 \ee
 and correspondingly $F=-2E^2$. For the line element (\ref{metric}) together with the electromagnetic tensor (\ref{F}), it can be seen that the relevant components of the equations of motion (\ref{GReq}) and (\ref{EMeq}) result in
\begin{eqnarray}
\frac{f'(r)}{2r}+\Lambda & = & 8\pi\left[L(F)+4E^{2}(r)L_{,F}\right],\label{eq:motion00}\\
E(r)L_{,F} & = & -\frac{q}{r},\label{eq:Elf}
\end{eqnarray}
where $q$ is an integration constant related the electric charge. Note that for the particular case of Maxwell theory, namely, $L(F)=F$, the electric field in (\ref{eq:Elf}) corresponds to the field of a point charge in (2 + 1)-dimensions $E(r)\sim 1/r$. It is convenient to express the derivative $L_{,F}$ as a function of $r$, and this is accomplished from Eq. (\ref{F}), leading to
\begin{equation}
E(r)L_{,F}=-\frac{L'(r)}{4E'(r)},
\end{equation}which compared with (\ref{eq:Elf}), results in
\begin{equation}
L'(r)=\frac{4q}{r}E'(r),\label{eq:Llinha}
\end{equation} where the prime ($'$) stands for the total derivative with respect to the radial coordinate $r$.

Thus, we can rewrite Einstein equation (\ref{eq:motion00}) equivalently as
\begin{equation}\label{flinha}
\frac{f'(r)}{2}+\Lambda r=8\pi\left[r L(r)-{4q}E(r)\right].
\end{equation}

In principle, we can integrate the above equation for all values of $r$. For the metric ansatz that we assume here, it can be verified that this condition is guaranteed if $f'(r)$, $rL(r)$, and $E(r)$ are free of singularities everywhere. Furthermore, to obtain regular black hole solutions, one must ensure that the curvature invariants, i.e., the Ricci scalar $R$ and Kretschmann scalar $K$ (built from the Riemann tensor) are singularity free. For this latter, we need the function $f(r)$ and its first and second derivatives to be free of singularities everywhere \cite{Cai:2006pq}.

The previous requirements impose constraints on the geometric side of Einstein equations, encoded in the general form of the metric function $f(r)$. The simplest regular black hole solution in (2 + 1)-dimensions is just the celebrated BTZ solution, with $f(r)=-m+r^{2}/l^{2}$. If the presence of matter fields is taken into account, then the metric function may include additional terms, and a standard ansatz to represent the metric function of a general regular black hole in (2 + 1)-dimensions is
\begin{equation}\label{efeerre}
f(r)=-m+\frac{r^{2}}{l^{2}}+k(r),
\end{equation}where $k(r)$ is a well-behaved function (continuous and differentiable) which because Eq. (\ref{flinha}) must obey the equation
\begin{equation}
\frac{k'(r)}{2r}=8\pi\left[L(r)-\frac{4q}{r}E(r)\right].\label{eq:k}
\end{equation}So, one can construct several regular black holes through a convenient choice of the Lagrangian $L(F)$ or the electric field $E(r)$ \cite{He:2017ujy}.

In addition to the geometric aspects discussed above, one must consider some conditions on the matter-energy content. To ensure a more physically acceptable origin for the solutions found, we still need to impose requirements on the matter fields, which in our case is represented by the Lagrangian of the nonlinear electrodynamics $L(F)$. It is customary to impose two requirements or conditions on $L(F)$: (1) the correspondence to the Maxwell theory in the weak field limit, namely $L(F)\rightarrow F$, for large $r$ and (2) the weak energy condition should be fulfilled, in our case requires $L(F)+4E^{2}L_{,F}\leqslant 0$.
The conditions described above are satisfied by many regular black hole solutions described in the literature.

As was pointed out before, one of our goals is to verify if it is possible to associate the matter source of the regular black holes solutions obtained by Estrada and Tello-Ortiz \cite{Estrada:2020tbz} and recently generalized by in Ref. \cite{Hendi:2022opt} to a nonlinear electromagnetic origin. For this purpose, we define an effective energy-momentum tensor $T^{\mu(eff)}_{\ \nu}=\mbox{diag}(-\rho,p_{r},p_{\theta})$ as follows,
\begin{equation}
T^{(eff)}_{\mu\nu} \equiv T_{\mu\nu}=g_{\mu\nu}L(F)-4L_{,F}F_{\mu\alpha}F_{\nu}^{~\alpha}.
\end{equation} In this way, we can carry out the following identification:
\begin{equation}
\rho(r)=-\left[L(r)-\frac{4q}{r}E(r)\right],\label{eq:rho1}
\end{equation}and with the help of Eq. (\ref{eq:Llinha}), we can express the Lagrangian $L(r)$ and the electric field $E(r)$ in terms of the energy density $\rho(r)$ as
\begin{equation}
L(r)=-\left[\rho(r)+r\rho'(r)\right],\label{eq:Lrho}
\end{equation}and
\begin{equation}
E(r)=-\frac{r^{2}}{4q}\rho'(r).\label{eq:Erho}
\end{equation}

According to Estrada and Tello-Ortiz \cite{Estrada:2020tbz}, an energy density compatible with the regular black hole solution in (2 + 1)-dimensions must meet the following requirements \footnote{These requirements were enunciated initially in work Eur.Phys.J.C 79 (2019) 3, 259 , e-Print: 1901.08724 [gr-qc].} (i) The energy density must be positive and continuously differentiable to avoid singularities; (ii) The energy density must have a single finite maximum at the origin; (iii) the energy density must be a decreasing radial function and must vanish at infinity. Moreover, for the case where the Lagrangian $L(r)$ and the electric field $E(r)$ are written in terms of the energy density $\rho(r)$, it is necessary to include the following additional requirement (iv) The energy density must be such that $r$ multiplied by the Lagrangian $rL(r)$ and the electric field $E(r)$ are free of singularities everywhere, as was mentioned after the equation \eqref{flinha}. In our case, for our election given for equations \eqref{eq:Lrho} and \eqref{eq:Erho}, it is direct to check that this latter is fulfilled due to condition (i).

Thus, they proposed the following functional form for the energy density:
\begin{equation}\label{StradaProf}
\rho(r)=\frac{m b^2}{\pi(b^2+r^{2})^{2}},
\end{equation}so that $\displaystyle \lim_{r\rightarrow\infty}\rho(r)=0,$ and $\rho(0)=m/\pi b^2$. Here, we are employing a slightly different definition, with $b$ being a positive constant with units of length $[b]=\ell$. The quasi-locality of the mass-energy \cite{Aoki:2020prb,Sorge:2020pdj} is assured, since $\int_0^{\infty} \rho(r) 2\pi r dr=m$ is finite . 

Thus, the Lagrangian $L(r)$ and the electric field $E(r)$ associated with this energy density are given by:
\begin{equation}
L(r)=-\frac{ m b^2(b^2-3r^{2})}{\pi(b^2+r^{2})^{3}},
\end{equation}and
\begin{equation}
E(r)=\frac{mb^2r^{3}}{\pi q(b^2+r^{2})^{3}}.
\end{equation}
We can easily see that the electric field does not tend to the expected asymptotic Maxwell-like behavior, $E(r)\sim 1/r$, in (2 + 1)-dimensions.

The functional form for the energy density profile given in Eq. (\ref{StradaProf}) can be generalized to arbitrary powers in $r$ and still preserve the quasi-locality properties (i)-(iv). Thus, we propose the following expression for the energy density:
\begin{equation}\label{rho1}
\rho(r)=\frac{A b^{n}}{(r^2+b^2)^{(n+3)/2}},
\end{equation}
with $n\geq 0$. Notice that, for $n=1$, we retrieve the form of Estrada and Tello-Ortiz energy density profile. The quasi-locality of mass-energy is guaranteed since we have
\begin{equation}
\int_0^{\infty} \rho(r) 2\pi r dr=m\Rightarrow A=\frac{(n+1)m b}{2\pi}.
\end{equation}
The electric field associated to this energy density, according to Eq. (\ref{eq:Erho}), is given by
\begin{equation}
E(r)=\frac{(n+1) (n+3) mb^{n+1}  r^3}{8\pi  q\left(b^2+r^2\right)^{(5+n)/2}}.
\end{equation}
Once more, we can observe that the field does not present an asymptotic Maxwell-like behavior, since in this limit $E(r)\sim r^{-2-n}$.

It is worth pointing out that another generalization for regular (2 + 1)-dimensional black holes was recently discussed in Ref. \cite{Hendi:2022opt}. However, our proposal contains it as a subset of sources for the possible regular black hole geometries.

At this point, we must raise the issue of the incompatibility between the mass-energy quasi-locality condition and the asymptotic Maxwell limit for the electric field, at least in (2 + 1)-dimensions. To one get the field behavior at the infinity as $E(r)\sim r^{-1}$, the required $\rho(r)$ must be represented by a series expansion, according to Eq. (\ref{eq:Erho}), in the form
\begin{equation}
\rho(r)=a_0r^{-2}+\sum_{n=0}^{\infty}b_n r^{n},
\end{equation}
where the sum converges in all domain. Since the mass is given by
\begin{equation}
m=\int_0^{\infty} 2 \pi r \rho(r)dr,
\end{equation}
we have that this quantity diverges. Therefore, we conclude that any energy density profile obeying the weak energy condition and the quasi-local energy requirement cannot be associated with an electrodynamics source since the Maxwell electrodynamic limit is not recovered for large $r$.

From the results obtained above, we can study the horizons associated with the regular black hole solution obtained from the quasi-localized energy density profile proposed by us. Thus, given that, by Eqs. (\ref{eq:k}) and (\ref{eq:rho1}), $k(r)$ is proportional to $\widetilde{m}(r)=\int 2\pi r\rho(r)dr$ plus an integration constant, the form of the metric coefficient should be, according to (\ref{efeerre}) and (\ref{rho1})
\begin{equation}\label{metrics}
f(r)=1-8m +r^2/\ell^2+ 8\widetilde{m}(r)=1-8m+\frac{r^2}{\ell^2}+\frac{8mb^{n+1}}{(r^2+b^2)^{(n+1)/2}},
\end{equation}
with the referred constant being $1-7m$, where we have chosen to get an anti-de Sitter space nearby the origin for the regular black holes and, at infinity, $\widetilde{m}(r) \to 0$, as expected from the quasi-locality of mass-energy. Since our solution is asymptotically similar to the BTZ metric, namely, $f(r)\approx 1-8m+r^2/\ell^2$, the curvature is entirely determined by $\Lambda<0$, and the spacetime is asymptotically AdS rather than asymptotically flat.

\begin{figure}[h!]
    \centering
            \includegraphics[width=0.55\textwidth]{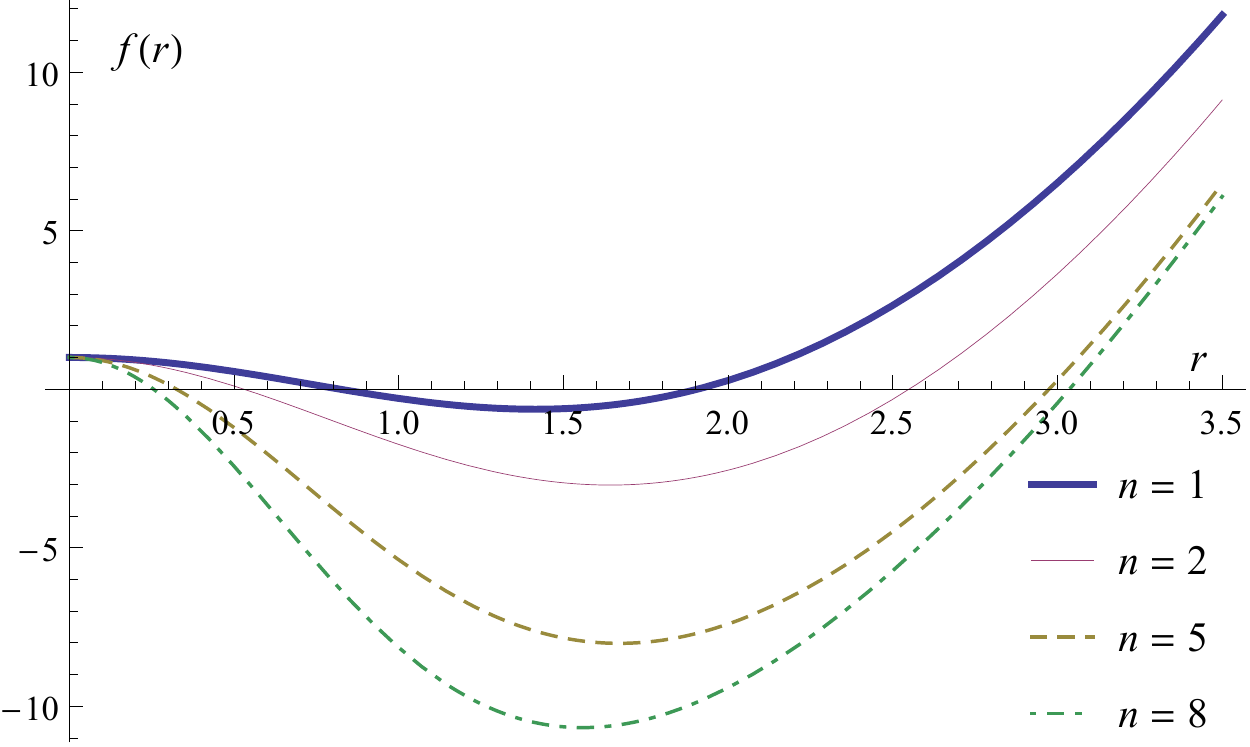}
        \caption{Metric coefficient for the regular black holes given from Eq. (\ref{metrics}), as a function of the radial coordinate, for some powers $n$, with $m=2.5$, $b=2.2$, and $\ell=0.7$ in Planck units.}
    \label{MCoef}
\end{figure}

Considering the Descartes rule, for $8 m\leq1$, there is no real roots for $f(r)=0$ and the black holes do not possess any horizons. For $8m>1$, these objects can have up to two horizons, as it is depicted in Fig. \ref{MCoef}. Notice that the radius of the outer (inner) horizon, $r_h$, is greater (smaller) the greater the power, $n$.

On the regularity of the obtained solution, we consider the curvature invariants, namely, the Ricci $R$ and Kretschmann $K$ scalars calculated below for the metric (\ref{metrics}):
\begin{eqnarray}
R	&=&-\frac{6}{\ell^{2}}+\frac{8mb^{n+1}(n+1)(3b^{2}-nr^{2})}{(r^{2}+b^{2})^{\frac{n+5}{2}}},\\
K	&=&8\left(\frac{1}{\ell^{2}}-\frac{4mb^{n+1}(n+1)}{(r^{2}+b^{2})^{\frac{n+3}{2}}}\right)^{2}\nonumber\\
	&& +\frac{4}{\ell^{4}(r^{2}+b^{2})^{n+5}}\left[(r^{2}+b^{2})^{\frac{n+5}{2}}+4(n+1)\ell^{2}mb^{n+1}\left((n+2)r^{2}-b^{2}\right)\right]^{2}.
	\end{eqnarray}
As we can see, Ricci and Kretschmann do not diverge at the origin $r=0$ or any other value of $r$ since we assume $b>0$. Thus, our solution (\ref{metrics}) represents a genuine class of singularity-free spacetime.

%%%%%%%%%%%%%%%%%%%%%%%%%%%%%%%%%%%%%%%%%%%%%%%%%%%%%%%%%%%%%%%%%%%%%%%%%%%%%%%%%%%%%%%%
\section{Thermodynamics of the three dimensional regular black holes\label{thermal}}

We start with the study of Hawking temperature expressed by $T_H=f'(r_h)/4 \pi$. The form of the metric function $f(r)$ (\ref{metrics}) allows us to determine this quantity analytically in terms of the horizon radius, and it is given by
\begin{equation}\label{HTemp}
T_H=\frac{r_h}{2\pi\ell^2} \Bigg\{1+\frac{(n+1) b^{n+1} \left(\ell^2+r_h^2\right)}{2\left(b^2+r_h^2\right) \left[b^{n+1}-\left(b^2+r_h^2\right)^{\frac{n+1}{2}}\right]}\Bigg\}.
\end{equation} For $b=0$, we retrieve the Hawking temperature of BTZ black hole.

\begin{figure}[!ht]
    \centering
    \begin{minipage}{0.44\linewidth}
        \centering
        \includegraphics[width=1.1\textwidth]{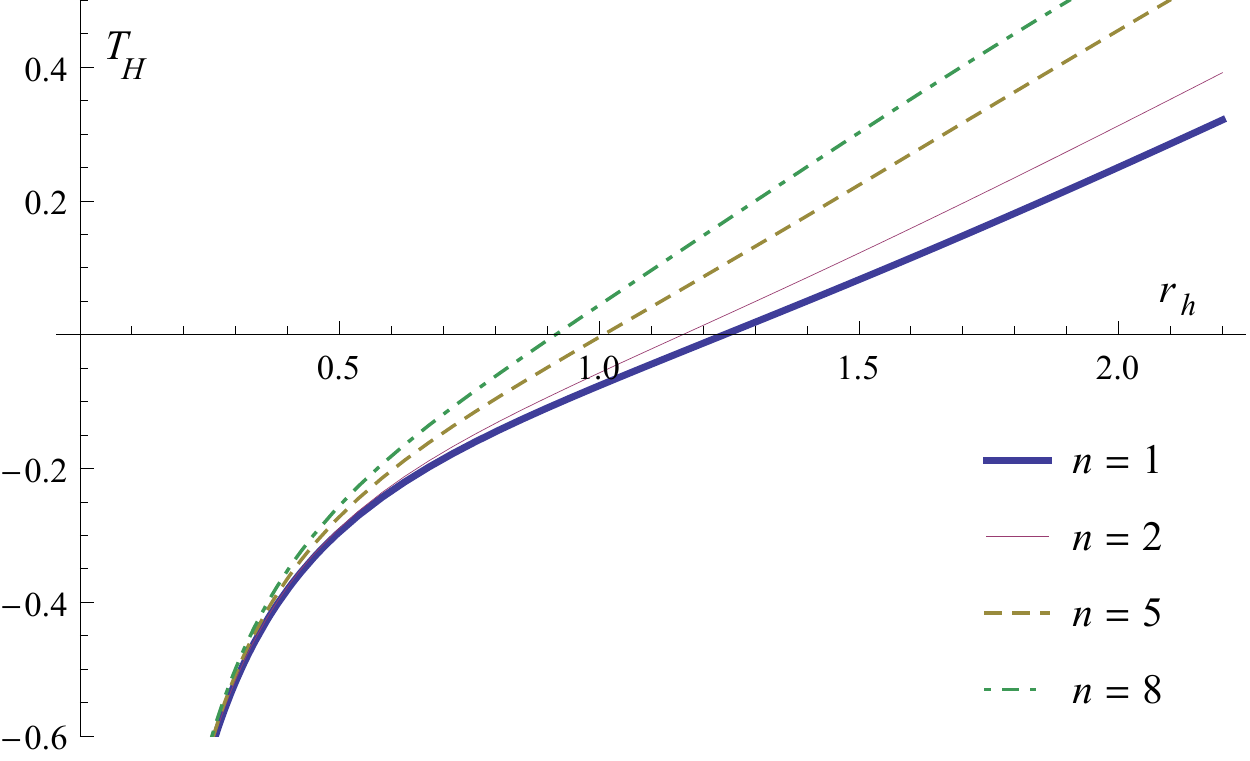}
            \end{minipage}\hfill
    \begin{minipage}{0.44\linewidth}
        \centering
        \includegraphics[width=1.1\textwidth]{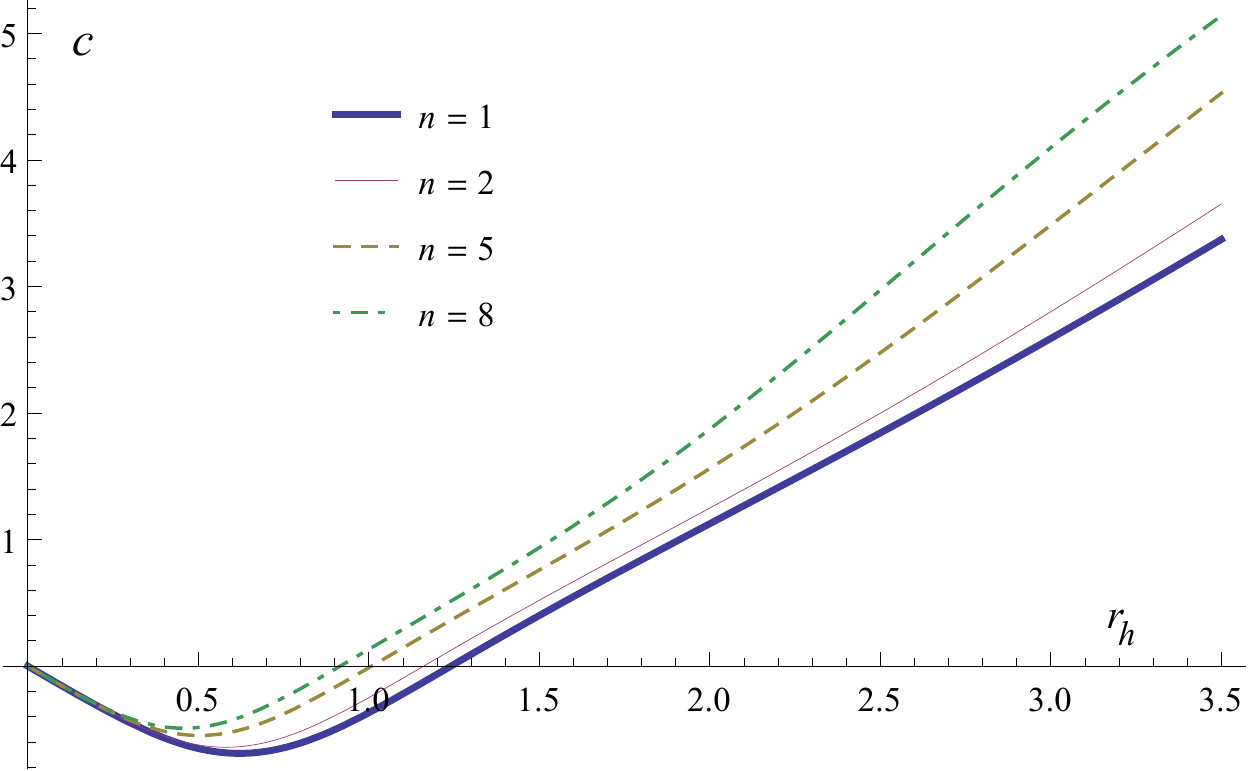}
                \end{minipage}
   \caption{Hawking temperature (left panel) and heat capacity (right panel) of the regular black holes under consideration, as functions of the event horizon radius  for some powers, $n$, with $b=2.2$ and $\ell=0.7$ in Planck units.}
    \label{HTempFig}
\end{figure}

We depict Eq. (\ref{HTemp}) in the left panel of Fig. \ref{HTempFig} for some black hole solutions, depending on $n$. Notice that the temperature is greater, the greater this power. From this figure, we can also infer that for the regular black holes under consideration, a critical horizon radius for which the temperature vanishes exists. That last statement means that when the black hole reaches this radius, it stops radiating, leaving a remnant behind. Note that the greater the power in $n$, the smaller the critical radius.

For the regular black hole solution in which $n=1$, we can exactly calculate this critical radius, which is given by
\begin{equation}
r_h^c=\sqrt{b\ell},
\end{equation}
and for any $n$, when $b=\ell$ we get
\begin{equation}
r_h^c=b\sqrt{4^{-\frac{1}{n+1}} (n+3)^{\frac{2}{n+1}}-1}.
\end{equation}

%%%%%%%%%%%%%%%%%%%%%%%%%%%%%%%%%%%%%%%%%%%%%%%%%%%%%%%%%%%%%%%%%%%%%%%%%%%%%%%%%%%
\subsection{A new propose for the structure of the first law of thermodynamics for regular black holes}

As mentioned in the introduction, the standard version of the first law, namely $dm=TdS-PdV$, leads to entropy and thermodynamics volume values that do not coincide with the usual definitions (for example, the entropy does not follow the area law in GR). So, to address this problem, we propose the following structure of the first law of thermodynamics for regular black holes
\begin{equation}\label{firstlaw}
dU=dm+dX=T_HdS-PdV,
\end{equation}
where $dX$ is a quantity that modifies the internal energy regarding the BTZ black hole, since we are working with a different class of regular three-dimensional black holes. 

It is worth mentioning that the equation \eqref{firstlaw} is locally defined at the horizon, $r_h$. Thus, the term $dU$ can be understood as a local definition of the variation of the internal energy at the horizon.

Following Ref. \cite{Kothawala:2007em}, we identify the radial pressure of the energy-momentum tensor as the thermodynamics pressure, which in our case is given by $P(r)=-\rho(r)=- m(n+1)b^{(n+1)}/[2\pi(r^2+b^2)^{(n+3)/2}]$, with $V=\pi r_h^2$. If we express $m$ as a function of $r_h$ from Eq. (\ref{metrics}), and considering $S=\pi r_h/2$ ({\it i.e.}, following area law \cite{Frodden:2012nu}), we can analytically find the quantity $X$, after integration of the Eq. (\ref{firstlaw}). Hence,
\begin{equation}
X=\frac{b^{n+1} \left(\ell^2+r_h^2\right)}{8 \ell^2 \left[b^{n+1}-\left(b^2+r_h^2\right)^{\frac{n+1}{2}}\right]},
\end{equation}
which corrects the internal energy of the regular black hole under consideration. For $n=1$, it assumes the simplest form
\begin{equation}
X=-\frac{b^2 \left(\ell^2+r_h^2\right)}{8 \ell^2 r_h^2}.
\end{equation}
On adding $m$ and $X$ in order to obtain the internal energy, we find the unexpected simple quantity
\begin{equation}\label{internalenergy}
U= \frac{1}{8} \left(1+\frac{r_h^2}{\ell^2}\right),
\end{equation}
for all $n$. Notice that the differential of this quantity is equal to the static BTZ black hole up to a multiplicative factor. Therefore, the internal energy is not identified with the quasi-localized mass of the regular black holes under inspection. This is because, as was above mentioned, the term $dU$ corresponds to a local definition of the variation of the internal energy at the horizon.

Identifying the cosmological constant as another component of the thermodynamics pressure, {\it i.e.}, $P_\Lambda=-\dfrac{\Lambda}{8\pi}$ \cite{Estrada:2020tbz}, from equation \eqref{internalenergy} we get to 
\begin{equation}
    U=P_\Lambda V + \frac{1}{8} \Rightarrow dU=P_\Lambda dV =dW_\Lambda
\end{equation}
where $dW_\Lambda$ can be understood as the work done by the thermodynamics pressure associated with the cosmological constant. It is worth mentioning that we are considering only variations along the horizon radius. In a future work could be considered variations along the cosmological constant, {\it i.e.}, in the extended phase space. Thus, we can say that considering the variation of the internal energy as the variation of the work done by the thermodynamics pressure associated with the cosmological constant in the first law of thermodynamics for regular black holes, the correct values of entropy and thermodynamics volume are obtained.

%%%%%%%%%%%%%%%%%%%%%%%%%%%%%%%%%%%%%%%%%%%%%%%%%%%%%%%%%%%%%%%%%%%%%%%%%%%%%%%%%%
\subsection{Stability analysis}

Since variations in volume and entropy are both functions of the horizon radius, $r_h$ in the parameter space, they are not mutually independent. Then, it is not possible to use the conventional definition of heat capacity given by the equation $(dQ/dT)_{v\sim r_+ =\mbox{fixed}}$ with $dQ=TdS$. This is only possible in the extended phase space, which could be studied in a future work. However, following \cite{Estrada:2019qsu}, it is possible to define the heat capacity in the form 
\begin{equation} \label{HC}
C=T_H\frac{dS}{dT_H}=T_H\frac{dS/dr_h}{dT_H/dr_h},
\end{equation} in order to analyze the evolution of the solution.

For $n=1$, we find
\begin{equation}
C=\frac{\pi  r \left(b^2+r^2\right) \left(r^4-b^2 L^2\right)}{2 \left[b^4 L^2+3 b^2 r^2 \left(L^2+r^2\right)+r^6\right]}.
\end{equation}
In the right panel of Fig. \ref{HTempFig}, we depict $C$ for some powers of $n$. Notice that the system presents thermodynamical local stability in the intervals where $r_h\geq r_h^c$, namely, until where the Hawking temperature vanishes, since there $C\geq 0$.

\begin{figure}[h!]
    \centering
            \includegraphics[width=0.55\textwidth]{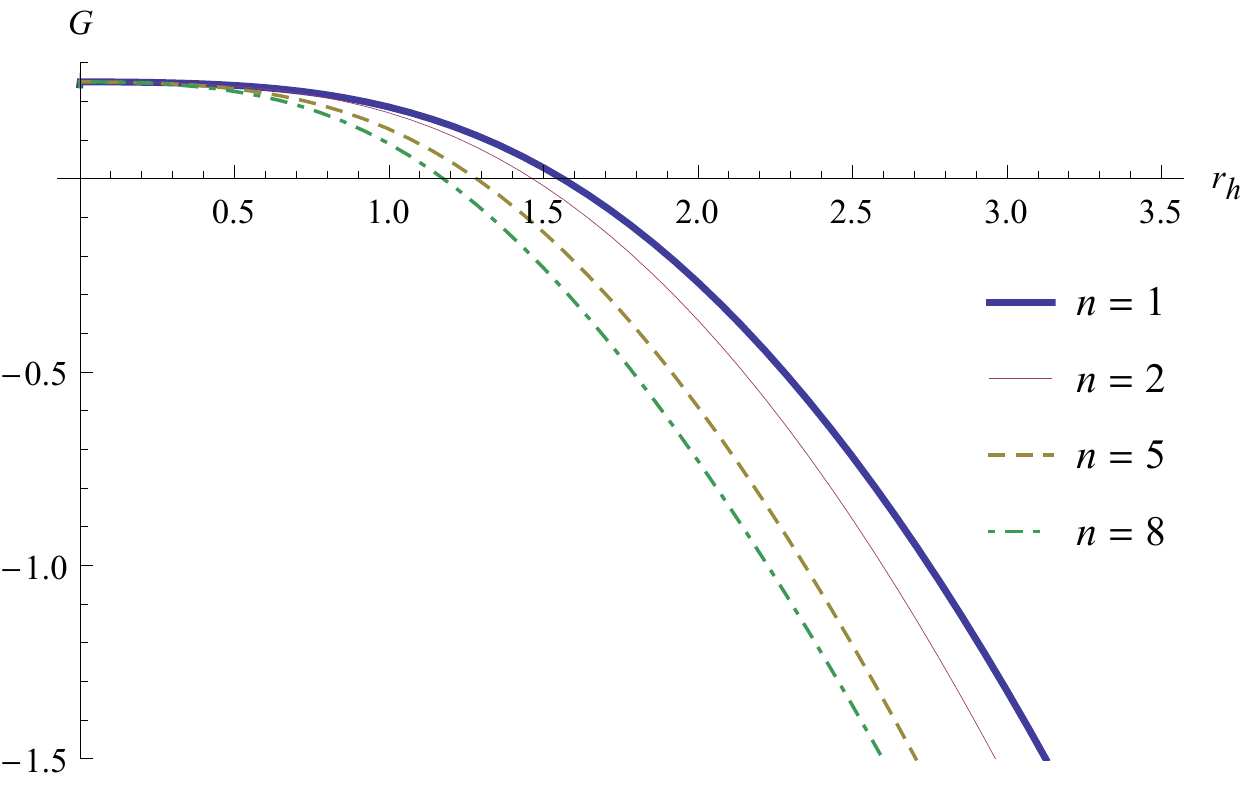}
        \caption{Gibbs free energy of the regular black holes, as a function of the event horizon radius, for some powers of $n$, with $b=2.2$, and $\ell=0.7$ in Planck units.}
    \label{Free}
    \end{figure}

Completing our thermodynamic analysis, we will calculate the Gibbs free energy, $G$, via $G=U-T_H S+PV$, and investigate the possible phase transitions. Thus, we get from Eqs. (\ref{HTemp}), (\ref{internalenergy}), and $S=\pi r_h/2$, that the Gibbs free energy is given by, for $n=1$
\begin{equation}\label{freeenergy}
G=\frac{2 b^2 \ell^2+\ell^2 r_h^2-r^4}{8 \ell^2 \left(b^2+r_h^2\right)}.
\end{equation}
From Fig. \ref{Free}, we can verify that the regular black holes realize phase transitions of zero order ({\it i.e.}, continuous) at a critical horizon radius, $r^{pt}_h$, where the Gibbs free energy vanishes.

Such a quantity goes thus from a globally thermodynamic stable region ($G<0$) to an unstable one ($G>0$) as it evaporates before reaching the critical horizon $r_h^c$ for which the remnant arises.

For $n=1$, we can exactly determinate the point of this phase transition:
\begin{equation}
r^{pt}_h=\sqrt{\frac{1}{2} \ell \left(\sqrt{8 b^2+\ell^2}+\ell\right)},
\end{equation}
which is always larger than $r_h^{c}=\sqrt{b\ell}$. We can also notice that the greater the free energy, the smaller is the power of $n$.

%%%%%%%%%%%%%%%%%%%%%%%%%%%%%%%%%%%%%%%%%%%%%%%%%%%%%%%%%%%%%%%%%%%%%%%%
\section{Conclusion \label{conclusion}}

We have obtained a new class of regular black hole solutions in (2 + 1)-dimensions whose source is a quasi-localized mass-energy distribution. These solutions depend on a characteristic distance, $b$, and an exponent parameter, $n$, with no horizons if the mass is smaller or equal to $m=1/8$ or with up to two horizons if $m>1/8$. The inner (outer) horizon is smaller (greater), with the greater power of $n$. The associated energy density satisfies the requirements described in Ref. \cite{Estrada:2020tbz}, namely, a positive profile, a single finite maximum at the origin, going to zero at infinity, and integrable in this interval, in what is known as quasi-locality of the mass-energy. Furthermore, it is satisfied a fourth requirement that we have included to write the electric field and the Lagrangian $L(F)$ as functions of the energy density. We show that, in general, such a source of matter cannot be associated with nonlinear electrodynamics since it does not lead to the Maxwell theory limit in the weak field limit for large $r$. In addition, we have shown that the referred conditions of quasi-locality to the energy density are incompatible with the expected electromagnetic asymptotic limit for the electric field, namely, $E \sim 1/r$. Our view is that the quasi-localized energy conditions are additional constraints on the matter sector. Similar to the fact that in (2 + 1)-dimensional spacetimes, general relativity does not reduce to Newtonian theory in the weak field limit. Ultimately, we say that our energy-matter source cannot come from nonlinear electrodynamics.

In the sequel, we studied the Hawking temperature for the regular black holes from the surface gravity on the event horizons. We found that such a temperature is greater, the greater the power, $n$, and that there exists a critical event horizon position, $r_h^c$, for which this temperature vanishes and the evaporation process halts, giving rise to a classical remnant mass.

Since the usual version of the first law of thermodynamics, $dm=TdS-PdV$, due to the presence of the matter fields, leads to incorrect values of entropy and thermodynamics volume for regular black holes, we have proposed a new version of the first law for regular black holes. So, we have included a potential $dX$ in the first law, which leads to a local definition of internal energy at the horizon, $dU=dm+dX$. We found the internal energy associated with the BTZ black hole, up to a constant, for all the found regular solutions. Furthermore, by identifying the cosmological constant as another component of the thermodynamics pressure, we could interpret the internal energy as the variation of the work done by the thermodynamics pressure associated with the cosmological constant $ dU=dW_\Lambda$. Remarkably, our form proposed for the first law allows us to obtain the correct values of entropy and thermodynamics volume. Hence, we found the heat capacity. It is positive for the horizon radius in which $r_h\geq r_h^c$, denoting thermodynamical local stability.

Finally, we determined the Gibbs free energy of the regular black holes under consideration, finding that a zero-order phase transition occurs at a critical event horizon larger than $r_h^c$, which separates a region globally stable from an unstable one, according to Fig. \ref{Free}.

%%%%%%%%%%%%%%%%%%%%%%%%%%%%%%%%

-------------------------------------------------------------------------

\section*{Acknowledgments}
\hspace{0.5cm} The authors thank the Coordena\c{c}\~{a}o de Aperfei\c{c}oamento de Pessoal de N\'{i}vel Superior (CAPES), Grant no 88887.621648/2021-00, and the Conselho Nacional de Desenvolvimento Cient\'{i}fico e Tecnol\'{o}gico (CNPq), Grants no
311732/2021-6 (RVM) and 308268/2021-6 (CRM), for financial support.

%%%%%%%%%%%%%%%%%%%%%%%%%%%%%%%%%%%%%%%%%%%%%%%%%%%%%%%%%%%%%%%%%%%%%%%%%%%%

\end{document}